# Paper 03

# Dynamic Simulation of Construction Machinery: Towards an Operator Model

Reno Filla [1], Allan Ericsson [1], and Jan-Ove Palmberg [2]

[1] VOLVO WHEEL LOADERS AB, ESKILSTUNA, SWEDEN
[2] DEPT. OF MECHANICAL ENGINEERING, LINKÖPING UNIVERSITY, SWEDEN

## Abstract

In dynamic simulation of complete wheel loaders, one interesting aspect, specific for the working task, is the momentary power distribution between drive train and hydraulics, which is balanced by the operator.

This paper presents the initial results to a simulation model of a human operator. Rather than letting the operator model follow a pre-defined path with control inputs at given points, it follows a collection of general rules that together describe the machine's working cycle in a generic way. The advantage of this is that the working task description and the operator model itself are independent of the machine's technical parameters. Complete sub-system characteristics can thus be changed without compromising the relevance and validity of the simulation. Ultimately, this can be used to assess a machine's total performance, fuel efficiency, and operability already in the concept phase of the product development process.

*Neo, sooner or later you're going to realize just as I did,*
*that there's a difference between knowing the path and walking the path.*

*(Morpheus in the film "The Matrix")*

# 1 Introduction

Originally spear-headed by large corporations of the automotive industry, dynamic simulation of complete vehicles is increasingly practiced in the development of off-road machinery. Handling and ride comfort are common simulation fields, as are performance and efficiency. As noted in [1], the challenge in simulating the complete system's dynamic behaviour is that in off-road machinery there are non-linear subsystems of various technical domains (drive train, hydraulics, electronics, mechanics, etc.), which are all tightly coupled. In the case of a wheel loader, the drive train and the hydraulics are parallel systems; both are competing for engine torque, which is in limited supply. Figure 1 shows how power is transferred through all relevant wheel loader subsystems, with the machine being used in the typical working task of loading gravel.

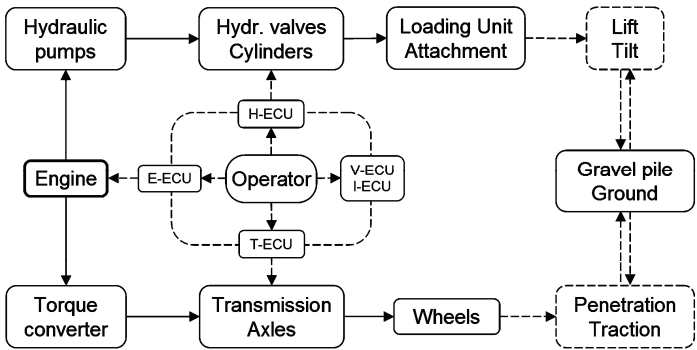

**Figure 1.** Simplified power transfer scheme of a wheel loader loading gravel

As described in [1], the momentary distribution of engine power to both parallel transfer paths is specific for the working task at hand, influenced by the environment, and controlled by the operator – who ultimately balances the complete system.

Therefore, in order to evaluate the complete system's performance, efficiency, and operability, the simulation must not be limited to the machine itself, but has to include operator, environment, and working task.

The results of a project aimed at simulating a wheel loader's environment, i.e. digging forces when working in gravel, have already been reported in [2]. Similar projects have also been documented in the literature, e.g. [3] and [4]. However, very little can be found on adding models of human operators to complete machine models, and using these to evaluate virtual prototypes in simulated working cycles. Mostly, current research is aimed at automating tasks. This paper will instead focus on the development of an operator model and a description of the working task, both sufficiently detailed to draw conclusions about a machine's total performance, efficiency, and operability.

## 2   Level of Detail

In this paper, "working task" will be defined as the summary of all descriptions of how the simulated machine shall be operated in its environment. The "operator model" describes how the machine shall be controlled to accomplish the working task.

In the simulation of complete machines interacting with their environment, the approach to modelling of operator and working task will very much depend on the questions that the simulation is to provide the answers to.

In Figure 2 (from right to left) it is shown that with a more detailed operator model, i.e. a more detailed model of a human being and their decision process, the description of the working task can be simpler, and restrictions on how the machine is to be operated are transferred from the working task description to the operator model. Instead of mere data, information is used – or even knowledge instead of mere information.

On the other hand, a less advanced operator model can to a certain degree be compensated for with more information or even significantly more data (Figure 2, from left to right). This will require less understanding of the total system but might still provide insightful simulation results, depending on the context (i.e. simulation goals).

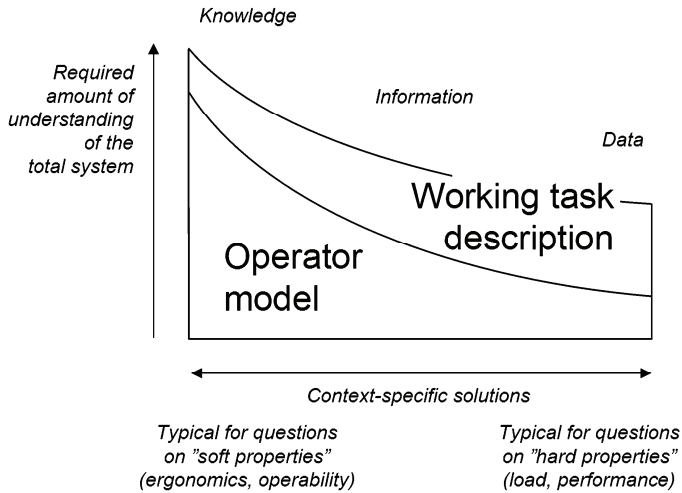

**Figure 2.** Relationship between operator model and working task description

When for example the aim of the simulation is to answer questions about mechanical loads on certain components of a wheel loader, the working task could consist of data for propeller shaft speed, steering wheel position, and current lengths of hydraulic lift and tilt cylinders – all recorded during a real loading cycle. The operator model could then be a simple PID controller that actuates the machine's input devices (throttle, brake, steering wheel, and hydraulic levers) in order to follow the recorded machine movement. Such a scenario would be located in the right hand area of the solution space in Figure 2. Performing an inverse simulation from the same set of working task data would abolish the need for an operator model completely.

At the opposite end (left side), one can find simulations for evaluation of human-related product properties like ergonomics or operability. Ideally, the operator as a sensitive, decision-making, and strategically planning human being is modelled in great detail. The working task could then be described in a very simple way, e.g. "load this gravel on a truck".

A decision for a specific level of detail can only be taken after careful consideration of the scope of the assignment. As always, it is important to specify what type of questions the simulation needs to give answers to – and what can be ignored.

## 3  Literature Review

In [5] Vogel gives a comprehensive overview of the driver behaviour models used in traffic simulation. She then combines some existing theories and develops her own, control theory-based framework. The existence of a "mental model" of the modelled system enables the driver to anticipate events and act preventively (instead of just mechanically reacting to target deviations). The author notes, "It is a very ambitious desire to provide a complete model of driver behaviour, and any such attempt will certainly provoke much criticism. On the other hand, not attempting at least to incorporate the possibility to model all aspects of driver behaviour can be criticized, too."

Another, much simpler, micro-level model for traffic simulation [6] uses a cellular automaton that incorporates mechanical restrictions (in the form of limited acceleration and braking capabilities) and some form of human behaviour (modelled as the driver's excessive response to local traffic conditions). Other examples are reported in [7] and [8]. Since simulation of traffic flow, just like road safety is a very active research area, many more papers, reports, and dissertations can be found.

Finding suitable strategies and control schemes for robotic excavation is another very active research area. Here, the focus is mostly on finding and following optimal bucket trajectories. Long Wu's approach to model bucket filling in [9] is motivated by his intention to develop an autonomous wheel loader. Among other things, he discusses bucket filling techniques and the shape of an optimal trajectory, in order to allow repetition in the following cycles. This latter aspect, giving autonomous machines tactical capabilities, has been previously covered by Singh [10] in great detail. In [11] he gives an update on the state of the art in 2002.

Another effort to automate the excavation process is reported in [12], which also features an extensive literature review, which also includes Hemami's early work [13]. Shi et al. recognize in [14] that "modeling the dynamics of tool/soil interactions is very difficult and computationally intensive, and thus not practical for real-time and online execution." Instead, they use fuzzy logic with very encouraging results.

With today's powerful computer hardware, Human-in-the-loop simulations are a feasible scenario, avoiding the necessity to model human behaviour (but of course sacrificing repeatability to a certain extent). One early example is reported in [15].

## 4  Development of the Wheel Loader Operator Model

Earlier projects within Volvo Wheel Loaders aimed at the development of a complete machine model, covering the simulation domains mechanics, drive train, hydraulics, and control system [16]. Having gained experience in projects involving multi-domain simulation of complete vehicles, Volvo Wheel Loaders wanted to extend these simulations to evaluate the potential operability of virtual prototypes, as well as total performance and fuel efficiency. These measures are heavily dependent on the way the human operator uses the machine. The aim was to develop an operator model, which should be as simple as possible with just the level of detail needed to reproduce relevant real-life phenomena.

Using a pre-recorded loading cycle as working task description and an operator model with a PID controller just following the cycle data (or even simpler: inverse simulation) would have worked for one specific machine. However, the slightest change in the setup (e.g. by altering torque converter characteristics) would either require new recorded data (which do not exist yet for virtual prototypes), or would inevitably produce erroneous results, when the working task description of the original machine setup was used. In [17] Zhang et al. correctly note: "The job is task-oriented, not reference-oriented. The operator is not explicitly following any speed or position reference when driving, steering and lifting. (…) The total productivity depends on how well the task is fulfilled, what the fuel economy is, and how long it takes to finish each cycle. Therefore, the performance and the efficiency of the human-machine interaction need to be maximized."

For such simulations, a human operator's ability to adapt to a new machine is needed to be reproduced by a model, preferably as simple as possible. Research in the areas of cellular automata and self-organization has shown that complexity can and often does emerge from quite simple, yet repetitive rules. In control theory, fuzzy logic is often used to derive a controller from the simple, yet vague rules a human operator would intuitively use for the same task. As stated earlier, Shi et al. have also reported on using fuzzy logic for robotic excavation [14].

These examples motivated the use of techniques similar to fuzzy sets and discrete events in the development of the operator model presented here. Since all subsystems of the machine as well as the gravel pile were modelled in the 3D Multi-Body Simulation package ADAMS (Figure 3), it was decided to also code the first version of the operator model in this simulation package.

Other simulation programs are surely better suited to handle this specific task, but using ADAMS proved to be feasible. Relationships between state of the system and human action were continuously described using cubic polynomial STEP functions (somewhat similar to fuzzy sets). Discrete states were saved using DIFF equations.

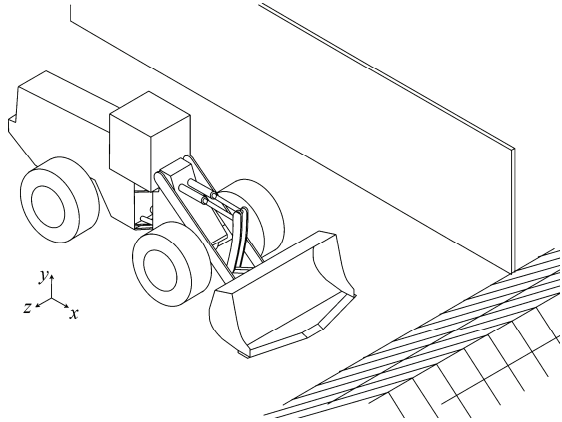

**Figure 3.** Models of the wheel loader and its environment, 3D view

All rules employed in the operator model were devised by studying and interviewing professional wheel loader operators. In [18] Gellersted documents a related study.

In the simulation, the operator model controls the machine model by engine throttle, lift and tilt lever, steering wheel, and brake only – just as a human operator does. Also, only signals that a human operator can sense are used in the model (excluding e.g. torque converter slip or hydraulic pump displacement). In its first version, the operator model has no tactical capabilities; it does not plan ahead of the current cycle by alternating the bucket fill location, like a human operator would do. The next sections will focus on the bucket filling phase itself, which is the most operator-dependent phase in a wheel loader's working cycle. All other phases have also been modelled, but will not be discussed in depth.

## 5 Bucket Filling

As mentioned, the literature reveals several different approaches to both modelling soil or granular material, and describing the optimal trajectory for a tool cutting through these materials. For this first version of the operator model, a less complex strategy was chosen: it was reasoned that the most efficient way to fill a bucket should be to move it upwards through the gravel pile on a velocity vector with a bearing $\delta$ that matches the pile's slope angle $\varepsilon$ (Figure 4). Of course, this will only be the case towards the end of the scooping, since the process starts with the bucket's cutting edge being parallel to the ground. Thereby, one specific bucket filling method is imitated, where the operator follows the slope of the gravel pile, instead of just forcing the bucket into the pile and tilting backwards.

At the same time, the bucket's cutting edge should remain at a certain angle of attack $\gamma$ relative to the bucket's velocity vector, and the bucket's bottom at a certain angle of clearance $\alpha$ relative to the gravel pile.

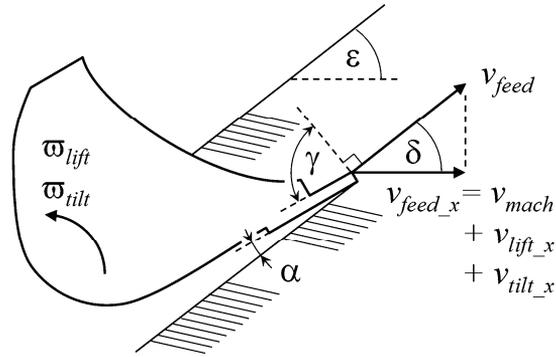

**Figure 4.** Bucket filling approach

The operator model needs to simultaneously control the machine speed (via the engine throttle) and bucket lift and tilt functions (via hydraulic levers) in order to satisfy these and a number of other requirements. In the simulation, the bucket filling phase starts when the bucket's cutting edge penetrates the gravel pile at a certain, controlled speed. This causes a down-shift into gear 1 and the repeated execution of the following rules (not necessarily in order):

- "Traction control 1": When relative wheel slip exceeds a certain value, the maximum throttle value will be ramped down. This decreases torque converter slip and thus reduces traction force.

- "Traction control 2": Above a certain limit for integrated relative wheel slip, the lift function is ramped up. This increases the load on the front wheels, which improves traction.

- "Bucket velocity vector control": Deviation of $\delta$ from $\varepsilon$ above a certain threshold will lead to a ramp-up of the maximum throttle value. This ensures that the bucket follows the gravel pile's slope.

- "Bucket attitude control": The tilt function is ramped up in order to maintain $\alpha$ and $\gamma$ as the lifting unit is raised and the machine is driven forward.

- "Exit trigger 1": Above a certain angle of the bucket relative to the gravel pile's slope, the tilt function will be fully activated until the bucket is completely tilted back.

- "Exit trigger 2": Above a certain angle of the lifting unit, the tilt function will be fully activated until the bucket is completely tilted back.

The results of rules governing the same operator input will either be multiplied or added to calculate the total input value. The bucket filling phase ends when the bucket has left the gravel pile, which will result in a gear shift into R2 (reverse) and full activation of the lift function.

## 6 Simulation Results

Figure 5 shows the results of one specific simulation, using the above rules for bucket filling:

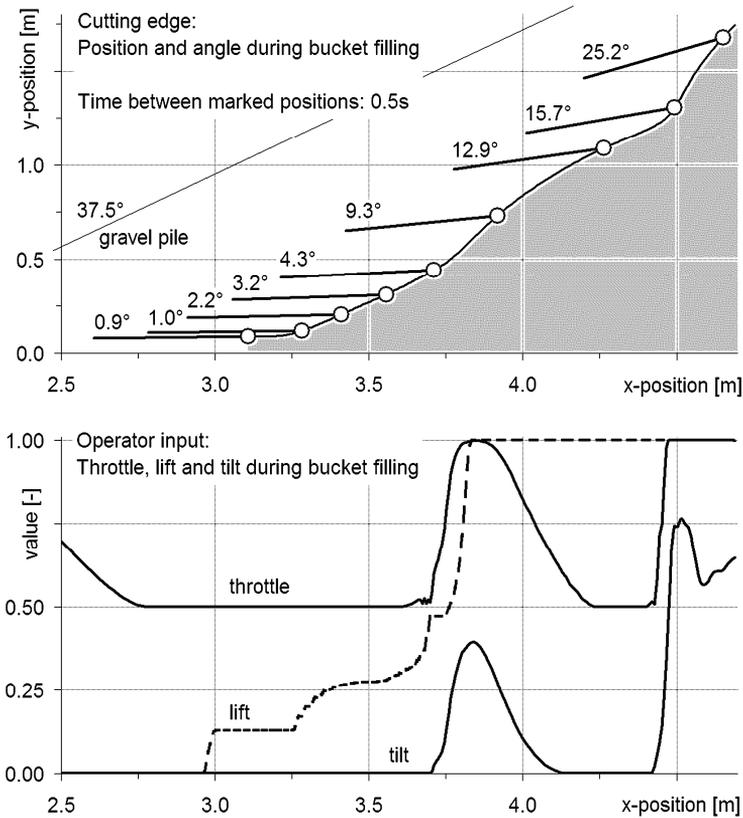

**Figure 5.** Bucket filling: operator input and simulation result

The upper diagram shows the x and y positions of the cutting edge and its global angle. The latter, visualized as sloped lines with attached values, is drawn each 0.5 seconds, which gives a feeling of the speed of the process. The lower diagram (which uses the same x axis) illustrates the operator input for engine throttle, lift, and tilt function (all calculated by merging the results of the previously defined rules).

Figure 6 shows the operator input during the complete loading cycle: engine throttle, brake, and steering wheel in the upper diagram, lift and tilt function in the lower one:

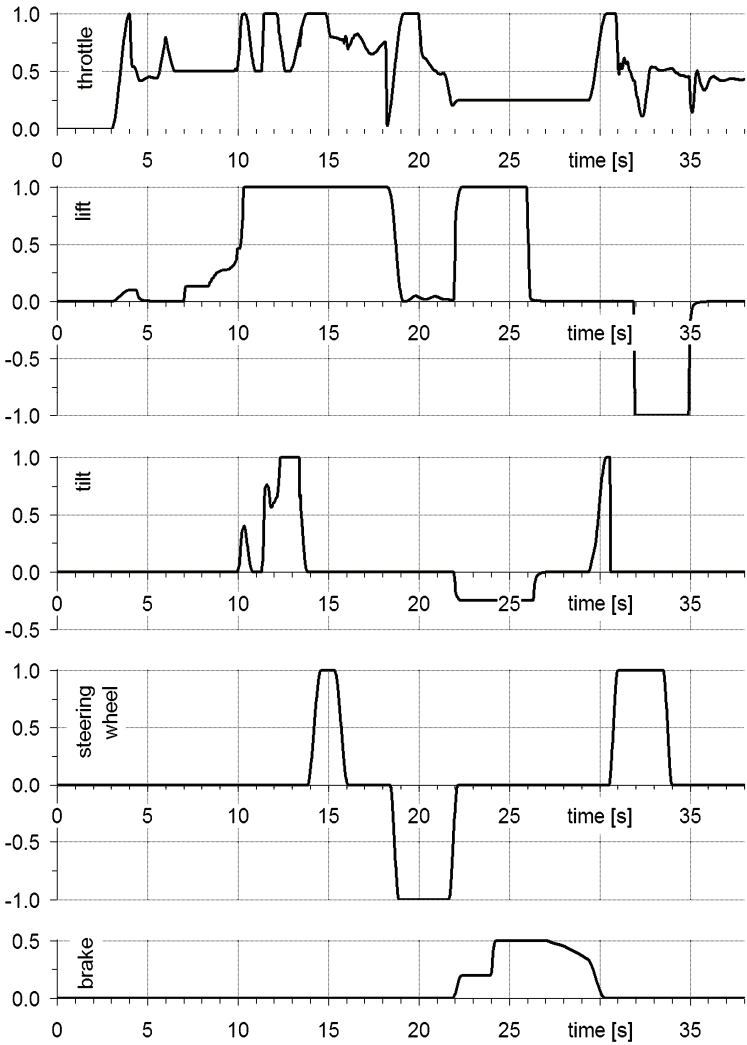

**Figure 6.** Complete loading cycle: operator input

As explained in the introduction, the motivation for the development of a sufficiently detailed operator model (and a description of the working task) was to be able to draw conclusions about a machine's total performance, efficiency, and operability by simulating virtual prototypes, rather than testing physical ones.

Of particular interest is how a wheel loader's engine power is being split up between drive train and hydraulics over a complete loading cycle. Figure 7 illustrates this for one of the conducted simulations:

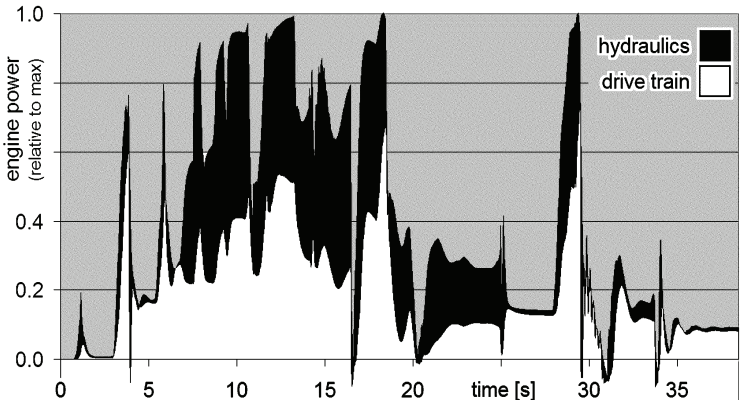

**Figure 7.** Complete loading cycle: power distribution to hydraulics and drive train (total engine power = top of the black area)

The engine's response and fuel consumption can vary dramatically depending on the specific combination of torque and speed that accomplished power. It is therefore important to analyze the engine's load duty (as shown in Figure 8):

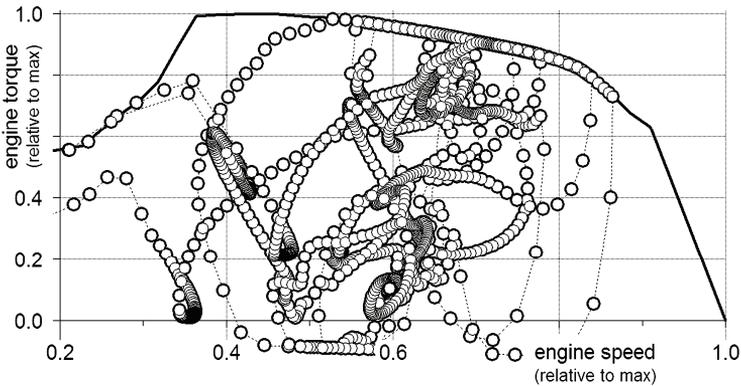

**Figure 8.** Complete loading cycle: engine load duty (normalized)

This pattern is remarkably similar to the results obtained in tests of physical prototypes, and indicates that the developed operator model can be useful. However, the goal is to be able to perform virtual tests of not yet physically realized machine configurations. It must therefore be proven that the operator model can adapt. The wheel loader model's torque converter has thus been changed to allow this. This component is known to have a vital impact on such important complete machine properties as performance and fuel efficiency. In the simulation, a "weaker" torque converter was chosen. In order to obtain the same traction force, a "weaker" converter requires higher slip between pump and turbine, which leads to higher engine speeds over a complete loading cycle.

A human operator compensates for this with higher throttle values. But this also affects the hydraulic system, which in turn requires compensation. Figure 9 shows that the operator model managed to adapt to the new machine characteristics by running the engine at higher speeds – just as a human operator would have done:

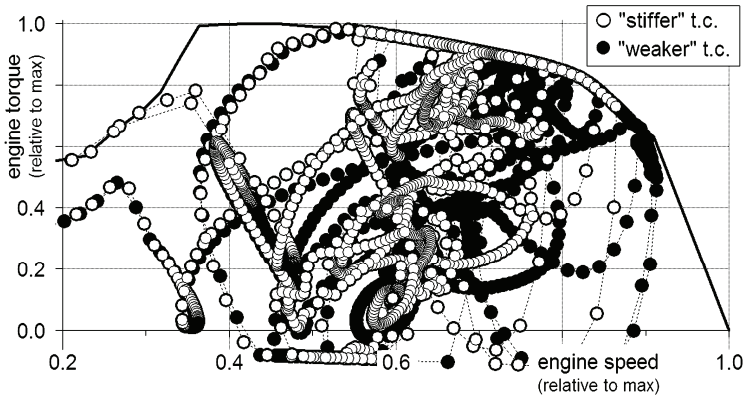

**Figure 9.** Complete loading cycle: engine load duty for machines equipped with different torque converters

The simulations conducted also replicate the fact that, in addition to higher engine speeds, a "weaker" torque converter also leads to higher fuel consumption and longer cycle times.

## 7   Discussion

In all conducted simulations, the operator model shows reasonably correct behaviour: the results for bucket filling, power distribution, engine load duty, and the ability to adapt to different torque converter characteristics all indicate that the model can be useful for testing total performance and fuel efficiency of virtual prototypes. However, the operator model in its first version clearly needs more development and validation.

In general, extracting controller rules from human operators through interviews and implementing them as min/max relationships or fuzzy sets is not a novelty. But this paper describes a new field of application with the ambition to connect to on-going work on quantification of a human operator's perception of a machine's operability. Applications in other areas, like active power distribution in hybrid vehicles, seem to be an interesting prospect.

Simulating operability is much harder than simulating total machine performance and fuel efficiency, as a generally agreed-upon definition of operability (especially for wheel loaders) is still missing. However, it should be possible to utilize the results of existing research into "mental workload", possibly connecting an operability measure with the operator's efforts to control the power distribution between hydraulics and drive train. If such a measure could be found, then simulation would be of great assistance in optimizing machine characteristics for example for maximum efficiency or robust operability (the latter both regarding component tolerances, varying environmental influences, but also different operator skills). One approach to quantification might be through the definition of "operator input dose" similar to vibration dose value in the assessment of whole-body vibration exposure.

Realizing the operator model as a set of equations in ADAMS proved to be possible, but cumbersome. For that type of problem, realization as a finite state machine in a discrete-event simulation package (in co-simulation with ADAMS) would have been better. This will be done in future work.

In the second chapter, the diametric relationship between operator model and working task description is discussed to a certain extent, but the following text seems to deal only with the operator model. This is because no distinction was made between the two in the code of the first prototype. The original idea was that with more knowledge, the working task description ("What to do?") transitions from demanding to describing, while the operator model ("How to do it?") changes from executing to planning. In this specific case, it has been found to be impractical to develop separate, yet linked models for these two. This could be so in other cases as well, turning the distinction of operator model and working task description more into a thought construct than a useful concept. On the other hand, similar concepts in other fields like computer science (object-oriented programming) or business management (project vs. line organizations) allow exceptions without dismissing the whole concept as totally irrelevant.

## 8   Conclusion

A first version of a rule-based operator model has been developed, that shows good potential for introducing "a human element" into dynamic simulation of complete wheel loaders. With this, more relevant answers can be obtained with regard to total machine performance and fuel efficiency in complete loading cycles. This can be used to significantly support the product development process by substituting many tests of physical prototypes with equivalent tests of virtual prototypes. However, using dynamic simulation to assess operability of complete machines still requires more work.

### Acknowledgements

The financial support of Volvo Wheel Loaders AB and PFF, the Swedish Program Board for Automotive Research, is hereby gratefully acknowledged.

Our sincere thanks are also due to the many people at Volvo and Linköpings Universitet without whose theoretical and practical support the work presented in this paper would not have been possible.

...Towards an Operator Model   15